\documentclass[preprint,authoryear]{elsarticle}

\usepackage{amsmath}%
\usepackage{listings}%
\usepackage[usenames,dvipsnames]{color}%

\begin{document}
\title{On the impact of dispersal asymmetry on metapopulation
  persistence}

\author[meg]{David Kleinhans\corref{dk}} \ead{david.kleinhans@gu.se}%
\author[met]{Per R.~Jonsson} \ead{per.jonsson@marecol.gu.se}

\cortext[dk]{Corresponding author}%
\address[meg]{University of Gothenburg, Department of Marine Ecology,
  Box 461, SE-405 30 G\"{o}teborg, Sweden}%
\address[met]{University of Gothenburg, Department of Marine Ecology,
  Tj\"arn\"o Marine Biological Laboratory, SE-452 96 Str\"omstad,
  Sweden}


\begin{abstract}
  Metapopulation theory for a long time has assumed dispersal to be
  symmetric, i.e.\ patches are connected through migrants dispersing
  bi-directionally without a preferred direction. However, for natural
  populations symmetry is often broken, e.g. for species in the marine
  environment dispersing through the transport of pelagic larvae with
  ocean currents. The few recent studies of asymmetric dispersal
  concluded, that asymmetry has a distinct negative impact on the
  persistence of metapopulations. Detailed analysis however revealed,
  that these previous studies might have been unable to properly
  disentangle the effect of symmetry from other potentially
  confounding properties of dispersal patterns.  We resolve this issue
  by systematically investigating the symmetry of dispersal patterns
  and its impact on metapopulation persistence. Our main analysis
  based on a metapopulation model equivalent to previous studies but
  now applied on regular dispersal patterns aims to isolate the effect
  of dispersal symmetry on metapopulation persistence. Our results
  suggest, that asymmetry in itself does not imply negative effects on
  metapopulation persistence. For this reason we recommend to
  investigate it in connection with other properties of dispersal
  instead of in isolation.

\end{abstract}

\maketitle

\section{Introduction}
Many species are structured in space with dispersal and migration
connecting local populations into metapopulations
\citep{Levins69,Hanski97}. The fundamental dynamics of metapopulations
are determined by local extinction, dispersal from the local
populations, and colonisation success leading to the establishment of
new sub-populations \citep{Levins69}. Metapopulation dynamics may
determine a range of ecological and evolutionary aspects including
population size \citep{Gyllenberg92}, persistence \citep{Roy05},
spatial distribution \citep{Roy08}, epidemic spread
\citep{McCallum02,Davis08}, gene flow \citep{Sultan02}, and local
adaptation \citep[e.g.][]{Hanski98,Joshi01}. Much interest has focused
on the effect of the spatial structure of metapopulations and how
local populations are connected through dispersal. Connectivity among
subpopulations is also increasingly emphasized in management and
conservation, e.g.\ to prevent fragmentation of landscapes
\citep{Crooks} and in the design of protected areas and nature
reserves \citep{vanTeeffelen06}.

Early models \citep[e.g.][]{Levins69,Hanski} assumed identical
dispersal probability among habitat patches. The initial focus of
spatially explicit metapopulation theory was to explore processes that
generate spatial patterns in homogeneous landscapes
\citep{Hanski02,Malchow}. Later, spatially explicit models were
designed to let dispersal probability be a function of patch size or
the distance between local habitat patches \citep{Hanski94,Hanski02}.
One aspect of dispersal that only has been implicitly included in
realistic models but not studied in isolation is when dispersal is
asymmetric. Asymmetric dispersal is expected for many metapopulations,
e.g.\ where dispersal is dominated by wind transport of pollen and
seeds \citep{Nathan01}, and for marine species with spores and larvae
transported by ocean currents \citep{Wares01}. Consequently, it is
important to understand how asymmetric dispersal may affect the
dynamics and persistence of metapopulations with potential
implications for the design of nature reserves. Some studies have
considered asymmetric dispersal
\citep[e.g.][]{Pulliam91,Kawecki02,Artzy-Randrup10} but have not
analysed effects on metapopulation viability.

In a recent contribution a conceptual model was developed to explore
the effects of dispersal asymmetry on metapopulation persistence
\citep{Vuilleumier06}. The viability of metapopulations was
investigated for different dispersal patterns randomly connecting
pairs of patches through either unidirectional or bidirectional
dispersal routes.  \citet{Vuilleumier06} concluded, that asymmetric
dispersal leads to a distinct increase in the extinction risk of
metapopulations. In a similar study \citet{Bode08} investigated
correlations between metapopulation viability and statistics of the
dispersal network; they also found that asymmetric dispersal links
resulted in higher extinction risk. Another very recently published
work investigates metapopulation viability for a selection of
asymmetric dispersal patterns and supports the findings of previous
works \citep{Vuilleumier10}.

The main objective with this study is to isolate the effect of
dispersal asymmetry from other properties of the metapopulation
network. When changing the degree of symmetry of dispersal networks
this generally may simultaneously influence the number of isolated
patches and other aspects of the network such as the balance of
dispersal in the individual patches \citep[see e.g. Figure 4
in][]{Vuilleumier06}. Since metapopulations are known to be sensitive
in particular to the density of the dispersal network
\citep{Barabasi04Review}, the existence of closed cycles of dispersal
\citep{Armsworth02}, and the hierarchy of dispersal in directed
networks \citep{Bode08,Artzy-Randrup10} these secondary implications
could confound any effect of asymmetric dispersal. We resolve the
problem by restricting our main analysis to \emph{regular} networks.

In this paper we in particular analyse the effect of asymmetric
dispersal on metapopulation persistence in more detail, with an
initial focus on regular dispersal networks.  We employ models of
synthetic dispersal patterns and demonstrate that asymmetric dispersal
per se may not lead to an increase in metapopulation extinction
risk. The significance of our results, their consequence for general
dispersal patterns and their relations to previous works are addressed
in detail in Section \ref{sec:discussion}.

\section{Material and Methods}
For ease of discussion we focus on the metapopulation model used in
previous approaches \citep{Vuilleumier06,Bode08,Vuilleumier10}. This
stochastic patch occupancy model connects a number of $N$ patches
through a complex dispersal matrix; the model is detailed in Section
\ref{sec:metapopulation-model-vuilleumier}. Within the scope of this
work the viability of metapopulations exposed to dispersal patterns
with different degree of symmetry is investigated. A consistent
definition of the degree of symmetry and details on the dispersal
patterns are provided in Sections \ref{sec:symmetry-def} and
\ref{sec:viab-metap-conn}.

\subsection{\label{sec:metapopulation-model-vuilleumier}Metapopulation model}
We consider a metapopulation consisting of $N$ patches of equal
quality, where, at a given time, each patch is either empty ($0$) or
populated ($1$).  Interactions of the patches are specified by means
of the $N\times N$ connectivity matrix $D$, where the elements
$d_{ij}\in\{0,1\}$ determine whether patch $j$ is connected to patch
$i$ ($d_{ij}=1$) or not ($d_{ij}=0$). For ease of discussion we
require $d_{ii}=0$ for all $i$ implying that patches are not connected
with themselves.

Building on previous works we used a stochastic discrete time model
for a metapopulation of $N$ patches and tested metapopulation
viability with respect to different connectivity matrices
\citep{Vuilleumier06,Bode08,Vuilleumier10}. The model, which is
attractive in its simplicity, implements dispersal through the
connectivity matrix $D$.  Initially all $N$ patches are populated. At
each time step two events occur in succession: first, populated
patches go extinct at per patch probability $e$. Subsequently, empty
patches can be colonised with probability $c$ by each incoming
dispersal connection from a populated patch. Newly populated patches
cannot give rise to colonisation of other patches at the same time
step they have been colonised.

In order to estimate the extinction risk of metapopulations the model
is iterated $T$ times. If any populated patch is left after the
$T^{\mbox{th}}$ iteration, the metapopulation is termed \emph{viable}
and \emph{extinct} otherwise. As \citet{Vuilleumier06} we chose the
parameters $e=0.5$ and $T=1,000$, and discuss the probability of
extinction of metapopulations consisting of $N=100$ patches as a
function of the colonisation probability $c$.

\subsection{\label{sec:symmetry-def}Symmetry of dispersal patterns}
For characterisation of the symmetry properties of dispersal patterns
the connectivity matrix $D$ is divided into its symmetric and
anti-symmetric contributions, $S$ and $A$, by defining the matrix
elements
\begin{subequations}\label{eq:def-a-s}
  \begin{eqnarray}
    s_{ij} & := & \min\left(d_{ij},d_{ji}\right)\\
    a_{ij} & := & d_{ij}-s_{ij}\quad.
  \end{eqnarray}
\end{subequations}
Based on these matrices the degree of symmetry $\gamma$ of dispersal
patterns is defined as the ratio of symmetric connections among all
connections:
\begin{equation}
  \gamma:=\frac{\sum_{i,j}s_{ij}}{\sum_{i,j}a_{ij}+s_{ij}}\quad.\label{eq:def-deg-asymm}
\end{equation}
Note that $1-\gamma$ is related to the asymmetry $Z$ discussed in
\citep{Bode08}.

By means of Equation \eqref{eq:def-deg-asymm}, the symmetry properties
of dispersal patterns are put on a firm footing: Dispersal patterns
are called \emph{symmetric} if $\gamma=1$ and \emph{anti-symmetric} if
$\gamma=0$.  Generally connectivity matrices $D$ with intermediate
$\gamma$ are neither symmetric nor anti-symmetric. We term them
\emph{asymmetric} if $\gamma<1$ corresponding to dispersal directed at
least to some degree.

\subsection{\label{sec:viab-metap-conn}Viability of metapopulations
  connected through regular dispersal patterns}

\begin{figure}
  \centering
  \includegraphics[width=\textwidth]{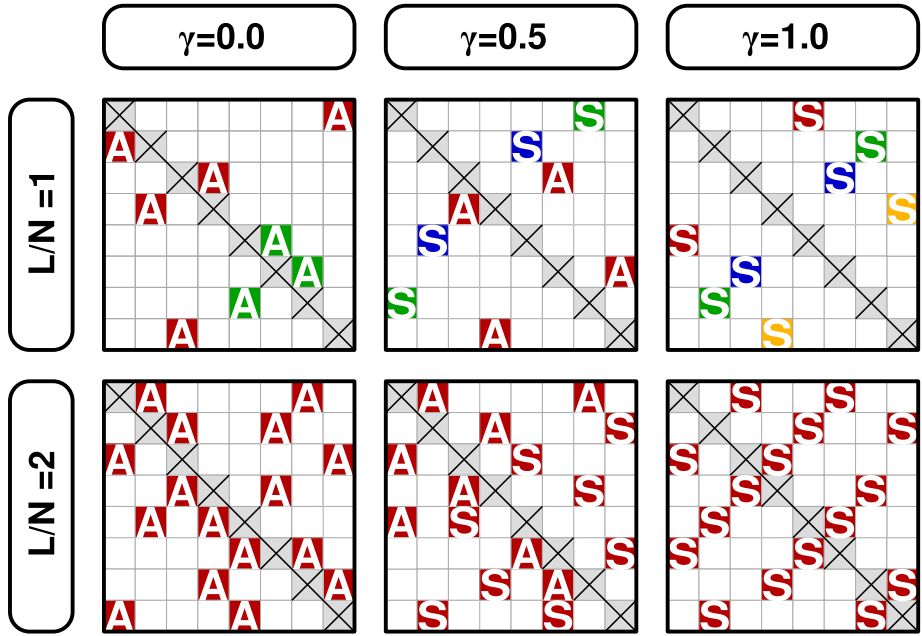}
  \caption{\label{fig:Algorithm-examples}Examples of connectivity
    matrices $D$ generated by the algorithm described in Section
    \ref{sec:viab-metap-conn} and in \ref{sec:algor-gener-balanc} for
    a reduced number of patches, $N=8$, and different combinations of
    $L/N$ and $\gamma$.  Only non-zero entries are printed
    explicitly. For reasons of clarity symmetric connections are
    denoted by 'S' and asymmetric connections by 'A'. The colours
    indicate separated closed cycles of dispersal that can be
    identified in the matrices. While the connectivity matrices with
    $L/N=1$ (upper row) are degenerate into $2$ ($\gamma=0.0$), $3$
    ($\gamma=0.5$), and $4$ ($\gamma=1.0$) clusters respectively, the
    clusters of all three matrices generated with $L/N=2$ (lower row)
    already extend to the entire metapopulation. In spite of the fact
    that the matrices displayed here are only \emph{examples} of
    randomly generated matrices, this trend is representative. For
    instance all simulations performed for $N=100$ and $L/N>2$
    resulted in dispersal matrices with a single cluster only. Note
    that our results are based on much larger metapopulations
    consisting of $N=100$ patches.}
\end{figure}
Previous works demonstrated that changes in the symmetry of dispersal
patterns in particular affect the local symmetry of migrant flow,
since asymmetry can result in \emph{donor}- and
\emph{recipient}-dominated patches not present in symmetric networks
\citep{Vuilleumier06}. In order to isolate the effect of the degree of
symmetry from these secondary effects, we focus on a specific set of
dispersal patterns: we restrict our main analysis to dispersal
patterns with the number of dispersal connections, $L$, being an
integer multiple of $N$ randomly distributed on the patches under the
constraint, that each patch obtains exactly $L/N$ in- and outgoing
connections with defined degree of symmetry. The random patterns
considered, hence, are \emph{regular} with the connections evenly
distributed to all patches available
\citep{Artzy-Randrup10,NetworkAnalysis}.  An algorithm efficiently
generating regular random dispersal patterns for small and
intermediate $L/N$ and arbitrary degrees of symmetry ($\gamma$) is
detailed in \ref{sec:algor-gener-balanc}.  Examples of random
connectivity matrices generated for $N=8$ and different combinations
of $L/N$ and $\gamma$ are exhibited in Fig.\
\ref{fig:Algorithm-examples}. Please regard that for the simulations
metapopulations consisting of $N=100$ are used resulting in
connectivity matrices of dimension $100\times100$ instead.

The regular dispersal patterns we use here restrict our analysis to
metapopulations with all patches connected at a fixed density
independent of the choice of $\gamma$. For $L/N>2$ the largest cluster
extends to the entire metapopulation independent from the degree of
dispersal symmetry resulting in irreducible connectivity matrices
\citep{Caswell01,Bode06}. For a detailed discussion of the impact of
regularity on our results we refer to Section
\ref{sec:discussion:regul}.

The viability of metapopulations exposed to these dispersal patterns
was tested in the following manner: a sample of $100$ dispersal
patterns connecting the $N=100$ patches was generated for each
combination of $L/N=1,\ldots,10$ and $10$ different values of
$\gamma$. For any of these patterns the viability of $10$ independent
realisations of metapopulations was tested for different values of the
colonisation probability $c$ according to the procedure outlined in
Section \ref{sec:metapopulation-model-vuilleumier}, resulting in a
statistics for a total of $1,000$ simulations on $100$ randomly
generated connectivity matrices for every choice of $L/N$, $\gamma$,
and $c$. For our main analysis we record the number of viable
metapopulations out of the $1,000$ simulations and prepare the results
for graphical analysis.  The sensitivity of this test procedure and
its interpretation with respect to the statistics of extinction times
is discussed in Section \ref{sec:discussion:interpret}.

\section{\label{sec:results}Results}
\begin{figure} \centering
  \includegraphics*[width=\textwidth]{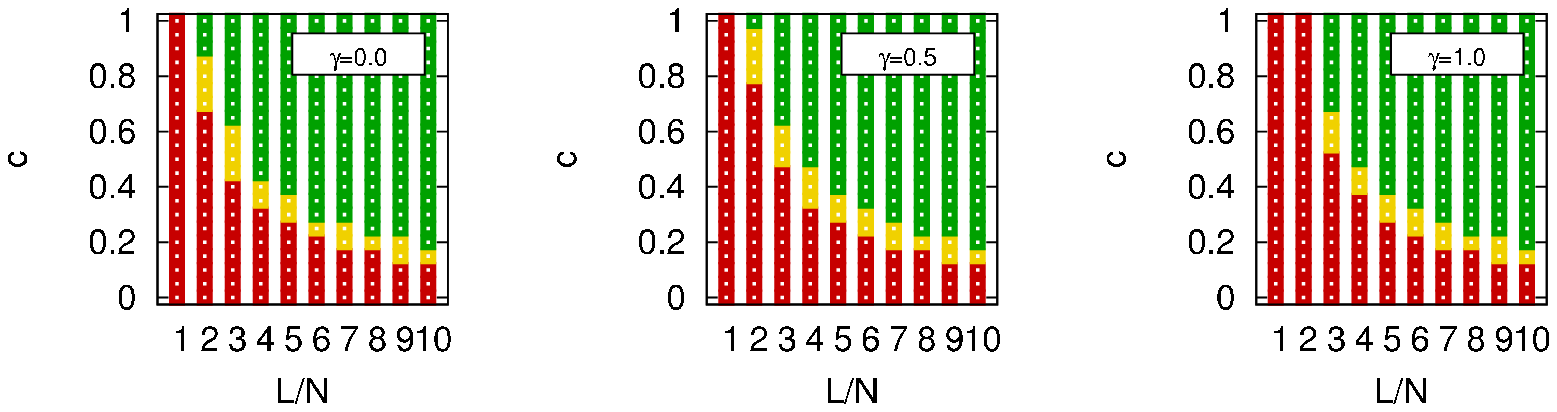}\\[4ex]
  \includegraphics*[width=\textwidth]{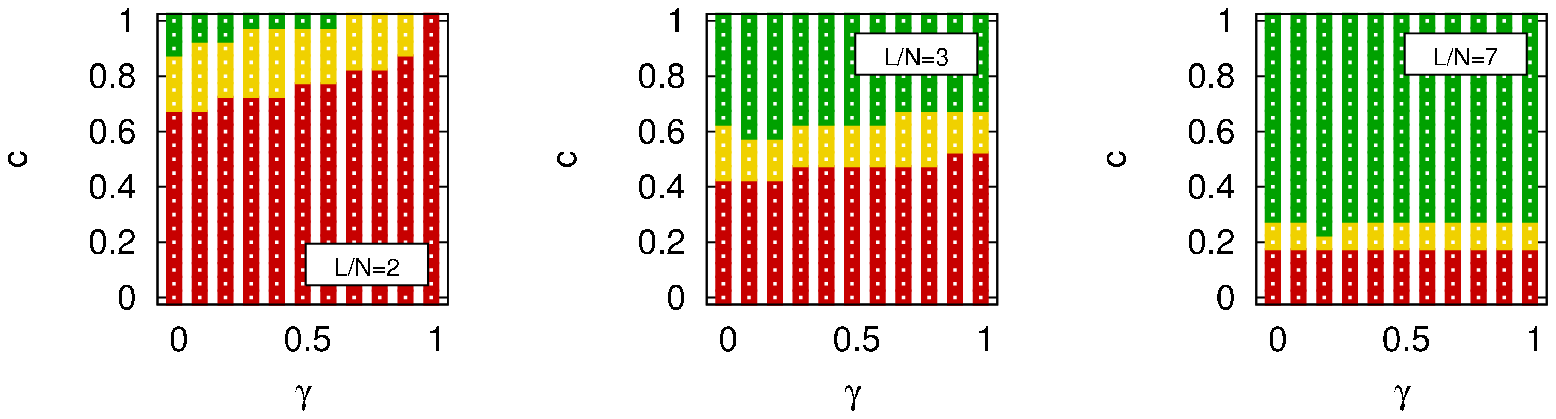}
  \caption{\label{fig:balanced}Results on the viability of
    metapopulations exposed to dispersal patterns with regular
    dispersal randomly generated by the algorithm described in Section
    \ref{sec:viab-metap-conn} and \ref{sec:algor-gener-balanc}. In the
    upper row the viability is plotted as a function of the effective
    number of connections per patch, $L/N$, and the colonisation
    probability $c$. At every combination of $L/N$ and $c$ the
    viability of $1,000$ different dispersal patterns has been
    investigated. Green and red squares indicate parameters, where
    either all $1,000$ patterns either were viable or not. The
    intermediate region where some of the patterns were viable and
    others were not is coloured yellow. The three panels present the
    results for different degrees of symmetry, increasing from
    $\gamma=0$ (anti-symmetric dispersal patterns) on the left to
    $\gamma=1$ (symmetric patterns) on the right hand side. In the
    lower row the simulation results are presented accordingly as a
    function of the symmetry $\gamma$ and the colonisation probability
    $c$ for three different number of connections per patch, $L/N=2,
    3$ and $5$. Only a vanishing impact of symmetry is observed for
    $L/N> 3$.}
\end{figure}
For each scenario $(L/N,\gamma,c)$ a total of $1,000$ simulations were
performed. For straightforward statistical evaluation of the viability
of metapopulations exposed to the respective conditions the simulation
results were divided into three different groups, which are colour
coded in the graphical presentation of the results: if all $1,000$
simulated metapopulations either went extinct or were viable the
scenario is coloured red or green, respectively. Otherwise, i.e.\ if
the number of extinct simulations out of $1,000$ is greater than $0$
but smaller than $1,000$, the scenario was coloured yellow.

The results are illustrated in Fig.\ \ref{fig:balanced}. The three
panels in the upper row show the viability of the metapopulation as a
function of the number of dispersal connections per patch, $L/N$, and
the colonisation probability $c$ for different values of $\gamma$:
anti-symmetric dispersal ($\gamma = 0.0$), asymmetric dispersal with
intermediate degree of symmetry ($\gamma = 0.5$), and symmetric
dispersal ($\gamma = 1.0$).  The lower panels of Fig.\
\ref{fig:balanced} contain the same results, but now analysed with
respect to the effect of the degree of symmetry, $\gamma$, for three
different values of $L/N$.  In fact, for $L/N > 3$ no statistically
significant impact of symmetry is observed.

\section{\label{sec:discussion}Discussion}
\subsection{\label{sec:discussion:interpret}Interpretation and
  significance of results}
First of all the results depicted in Fig.\ \ref{fig:balanced} suggest
that the impact of the degree of symmetry on metapopulation viability
decreases with increasing $L/N$. Already at $L/N>3$ no statistical
significant impact of the degree of symmetry, i.e.\ no systematic
differences depending on the degree of symmetry, can be detected on
the basis of the scenarios and the statistical evaluation chosen.

At a small number of dispersal connections per patch ($L/N=1,2$)
metapopulation viability is significantly reduced for more symmetric
dispersal (Fig.\ \ref{fig:balanced}, lower panels). The reason for
this effect straightforwardly can be understood from considerations
concerning the structure of the underlying dispersal patterns: Let us
first focus on patterns with $L/N=1$. In this case a metapopulation
with a symmetric dispersal pattern necessarily consists of a number of
patches only pairwisely connected through dispersal (Figure
\ref{fig:Algorithm-examples}). The largest closed dispersal cycle
(synonymous to the giant component of the dispersal network
\citep{Berchenko09}), hence, involves only two patches. For the
particular metapopulation model applied a lower bound for the
extinction probability of a pair of patches per time step is $e^2$. On
the contrary the mean size of the largest closed dispersal cycle
estimated from the $100$ dispersal patterns generated for the same
conditions but antisymmetric dispersal ($\gamma=0$) was $62.7$. For
$L/N=2$ the mean size of the largest cycles was $77.5$ for the
symmetric dispersal matrices generated, whereas for the asymmetric
case all dispersal matrices already extended to the entire
metapopulation (i.e.\ their mean size was $100$). Hence we are faced
with a percolation problem on random graphs \citep{Callaway00}, where
the percolation threshold depends on the symmetry properties of the
dispersal pattern. Analysis of the eigenvalues of associated state
transition matrices reveals, that the mean time to extinction of a set
of patches participating in a closed cycle of dispersal increases with
the size of the cycle. For this reason differences in viability at
small $L/N$ are attributed to hierarchical differences of the
generated matrices at only a few number of connections, namely $L/N\le
3$. This density is much smaller then relevant cases discussed e.g.\
in \citep{Vuilleumier06} as will be discussed in more detail in
Section \ref{sec:discussion:prev}.

\begin{figure}
  \centering
  \includegraphics[width=.6\textwidth]{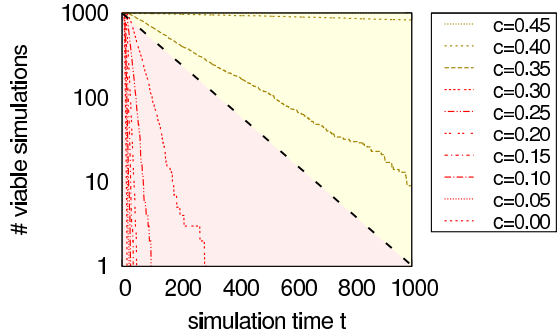}
  \caption{\label{fig:E50-itime}Extinction statistics for the
    metapopulations with different values of the colonisation
    probability $c$ connected through dispersal matrices with
    $\gamma=0.5$ and $L/N=4$. The individual lines indicate the number
    of non-extinct simulations (out of $1,000$) as a function of the
    simulation time. The dashed line corresponds to the upper bound
    for the expectation value of the number of extinct simulation for
    cases where all simulations went extinct, $1,000\exp(-6.9\times
    10^{-3} t)$, as derived in the manuscript text. From the figure it
    becomes evident, that the number of non-extinct simulations after
    an initial relaxation phase indeed decreases exponentially in time
    (i.e.\ linear in this logarithmic plot). The upper bound
    approaches $1/M$ with $t\to T$, which is a general result for
    sufficiently large $M$ and $T$ as a Taylor expansion of expression
    \eqref{eq:sig-inflection} shows. For this reason the boundary line
    indeed exhibits the border between the cases marked red and yellow
    in Figure \ref{fig:balanced}.}
\end{figure}

How meaningful is the statistical evaluation of the results with
respect to the effect of the symmetry of dispersal patterns on
expected extinction times of metapopulations? In order to approach
this question we aim to derive lower and upper bounds for extinction
times in the red and green regions of the figures, which then help to
evaluate the graphical presentation of the results in more detail. If
we disregard the initial time period of relaxation of the
metapopulation to a quasistationary state, we can assume that the
statistics of extinction times is exponentially distributed. This
exponential distribution complies with a constant risk of
metapopulation extinction per time step, which we call $r$. The
chance, that a metapopulation has not gone extinct after $T$ time
steps then is $(1-r)^T$. For every combination of parameters we
perform $M$ simulations with $M=1,000$ in our case\footnote{For
  reasons of clarity we here assume that simulations are independent
  of one another although in each case $10$ of them share the same
  dispersal patterns. This assumption, however, is not expected to be
  too extensive as the investigation of the replicate statistics at
  the end of Section \ref{sec:discussion:interpret} suggests.}. It is
then straightforward to calculate the probability $P(M|r)$ that all
$M$ simulations are viable,
\begin{equation}
  \label{eq:pMr}
  P(M|r)=(1-r)^{M T}\quad.
\end{equation}
Accordingly the chance that a simulation goes extinct during the $T$
simulation steps is $1-(1-r)^T$, resulting in the probability $P(0|r)$
of observing $0$ viable simulations of
\begin{equation}
  \label{eq:p0r}
  P(0|r)=\left(1-(1-r)^{T}\right)^M\quad.
\end{equation}
More interesting, however, would be the expressions $P(r|M)$ and
$P(r|0)$, the probability distributions of the metapopulation
extinction risk $r$ given the fact that either all or none of the
simulations are viable. These expressions straightforwardly can be
calculated using Bayes' theorem. Using uniform prior distributions we
obtain
\begin{eqnarray}
  \label{eq:prM}
  P(r|M)&=&\left(\int_0^1dr' (1-r')^{M T}\right)^{-1}(1-r)^{M T}\quad
  \mbox{and}\\
  \label{eq:pr0}
  P(r|0)&=&\left[\int_0^1dr' \left(1-(1-r')^{T}\right)^M\right]^{-1}\left(1-(1-r)^{T}\right)^M\quad.
\end{eqnarray}
Using a maximum likelihood approach confidence intervals for $r$ can
be calculated. Applying a confidence level of $95\%$ the upper bound
for $r$ in cases where all simulations are viable is $5.1\times
10^{-8}$. As a lower bound for $r$ for cases where all simulations
went extinct we obtain $0.057$. Since the latter result strongly
depends on the prior distribution we instead use the inflection point
of the sigmoid function \eqref{eq:pr0} at
\begin{equation}
  \label{eq:sig-inflection}
  1-\left[(T-1)/(MT-1)\right]^{1/T}  
\end{equation}
as a more conservative estimate, which for the case of our simulations
is located at approximately $6.9\times 10^{-3}$. 
The inverse of $r$ corresponds to the mean time to extinction. From
our considerations we, hence, expect the mean time to extinction for
the scenarios marked by red squares in Figure \ref{fig:balanced} to be
below $6.9^{-1}\times 10^3\approx 145$ and the respective value for
the conditions marked green to be in the order of $2\times 10^7$ or
larger. Intermediate values are expected for the conditions marked
yellow in the individual plots. Figure \ref{fig:E50-itime}
demonstrates, that assumptions we needed to make seem to hold and that
the estimates indeed reflect the underlying extinction statistics to a
great extent.

Obviously the classification of the conditions by the three scenarios
to a meaningful extent reflects the extinction risks of the
metapopulation in a sense, that Figure \ref{fig:balanced} succeeds to
highlight the main results. From the bounds for the mean extinction
times to extinction derived above for the respective classes we can
conclude that metapopulations in the red regions almost surely go
extinct within a short time, whereas metapopulations in the green
regions are likely to be persistent. The yellow region decreases in
range with increasing $L/N$. That is, the transition between
threatened and persistent metapopulation sharpens with increasing
$L/N$.

\begin{figure}
  \centering
  \includegraphics[width=.5\textwidth]{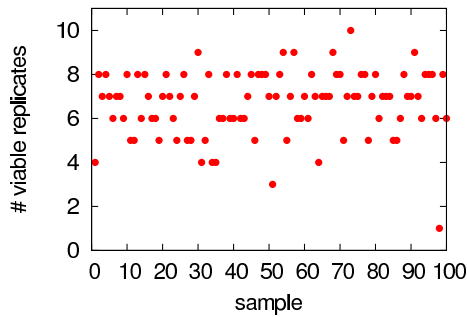}
  \caption{\label{fig:E50-replicate}Extreme example of the variation
    in the number of viable replicates between the different samples
    (here: $\gamma=0.3$, $L/N=10$, $c=0.15$). In particular sample
    $98$ deviates strongly from the general mean. Since we can assume
    that the main source of variations is the stochastic simulation
    procedure rather than qualitative differences between the random
    dispersal patterns relevant for the present study, we do not
    investigate the within-sample variations further within the scope
    of this work.}
\end{figure}

The $10$ replicate simulations performed for each parameter set and
each dispersal pattern in addition allow to investigate and to discuss
the variability within the sample of $100$ dispersal patterns. In the
regions marked red and green by definition all samples show the same
behaviour. Detailed analysis of the yellow regions shows only very few
cases of large variability of the number of extinct replicates between
the samples. One example of rather high variability is depicted in
Figure \ref{fig:E50-replicate}. Overall the differences between the
random dispersal patterns generated for each scenario do not seem to
be relevant for the present study, which is probably due to the
decision of using regular dispersal patterns.

\subsection{\label{sec:discussion:regul}Impact of regularity on the
  results}
So far we focused on regular dispersal patterns. This approach made it
possible to investigate the impact of the degree of symmetry of
connectivity matrices on metapopulation viability independently from
other possibly confounding effects, which is important in order to
assess the role of dispersal symmetry for metapopulations.  Our
results on regular dispersal patterns show a remarkably low effect of
symmetry ($\gamma$) on the viability of metapopulations at
intermediate and high density of dispersal paths, $L/N$. At low $L/N$
symmetric dispersal even results in a slightly negative effects on the
viability. How do these results relate to the more general case where
the dispersal network is not regular?

\begin{figure} \centering
  \includegraphics*[width=\textwidth]{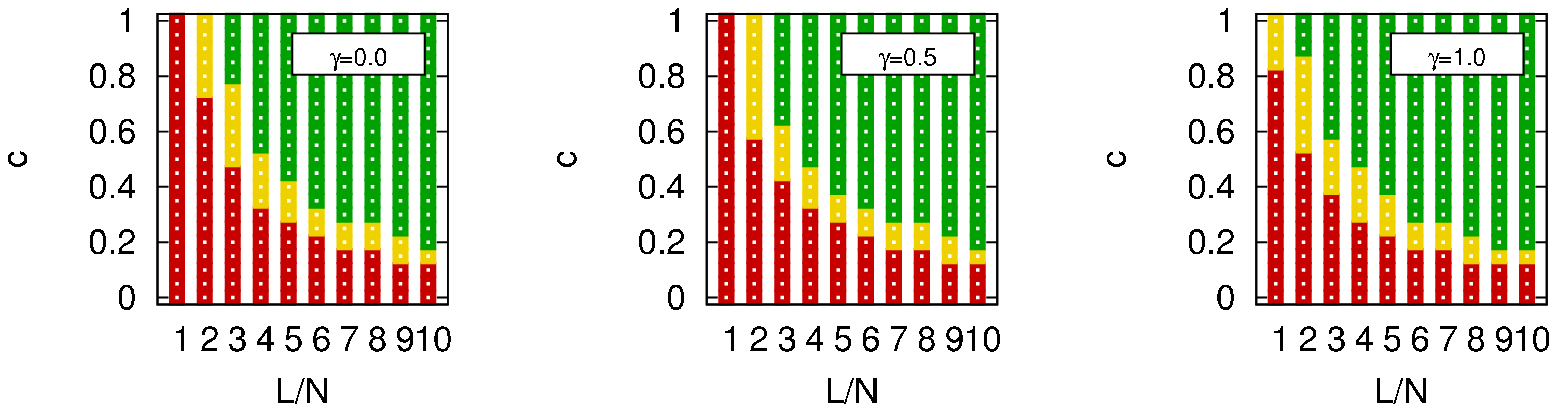}\\[4ex]
  \includegraphics*[width=\textwidth]{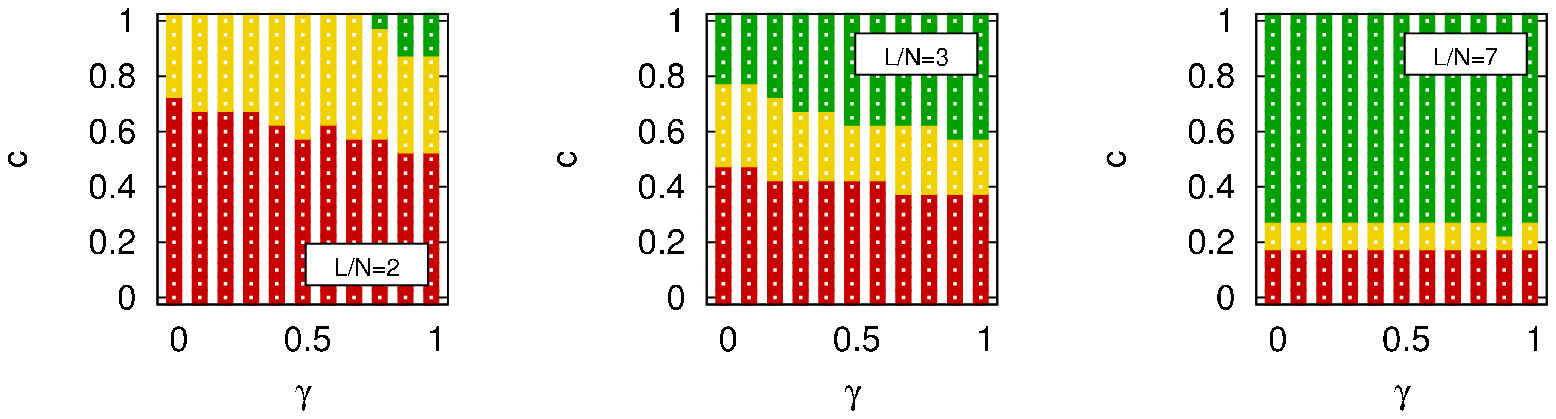}
  \caption{\label{fig:nonreg}Results on the viability of
    metapopulations exposed to general dispersal patterns randomly
    generated by modification of the algorithm described in
    \ref{sec:algor-gener-balanc}. The analysis and graphical
    presentation of the simulation results is accordant to the
    procedure described in the caption of Figure
    \ref{fig:balanced}. Please note that $L/N$ now specifies the mean
    number of connections per patch, while the actual number of
    dispersal links now can vary between patches.}
\end{figure}

In order to follow up this question we repeated the simulations
accordingly, but now without the constraint of having regular
dispersal networks. Technically this was implemented by skipping steps
4c and 4d of the pattern generation algorithm detailed in
\ref{sec:algor-gener-balanc}, which then controls for the desired
degree of symmetry only. The parameter $L/N$ now should be understood
in a statistical sense, such that $L$ dispersal connections randomly
were distributed between the $N$ patches resulting in a mean density
of $L/N$ connections per patch. The results are depicted in Figure
\ref{fig:nonreg}. Interestingly, the minor effect of symmetry at low
density of dispersal connections now shifts to a slight advantage for
metapopulations with a symmetric dispersal pattern. From $L/N\ge 7$ no
significant differences with respect to the simulation results based
on regular dispersal patterns (Figure \ref{fig:balanced}) are
observed.

In non-regular dispersal patterns the existence of isolated patches
not participating in dispersal has an impact on the effective density
of dispersal connections in the metapopulations \citep[see
also][]{Bode08}. Moreover, in the case of asymmetric dispersal there
exist patches that either only receive or only provide migrants, i.e.\
\emph{sinks} or \emph{sources}, and that cannot actively take part in
the metapopulation dynamics \citep{Artzy-Randrup10}. Since both of
these effects are most distinct at small densities of the random
dispersal networks, we assume that these differences basically drive
the minor differences at low $L/N$ between our results on regular and
the general case of random dispersal. Arguments for \emph{not}
assigning this effects to asymmetry in dispersal but to examine them
separately are made in Section \ref{sec:conclusions}.

\subsection{\label{sec:discussion:prev}Relation to previous works}
In general our results suggest essentially no direct negative effect
of asymmetric dispersal on metapopulation viability at intermediate
and high densities of the dispersal network, at least as far as the
stochastic patch occupancy model applied in this work is
concerned. This is in contrast to the findings in
\citep{Vuilleumier06} where it was concluded that extinction risk
significantly increased when dispersal became asymmetric.  The
analysis in \citep{Vuilleumier06} is not restricted to cases with
regular dispersal only, although the relaxation of regular dispersal
is not not sufficient to explain the qualitative differences in the
results as shown in the previous section.

The description of the random patterns investigated in
\citep{Vuilleumier06} does not provide all information necessary for
an in-depth comparison with our results. In \citep{Vuilleumier06} the
number of dispersal connections was chosen randomly for each of the
$2,000$ metapopulations. Additional information provided on two
particular patterns suggest that the densities are comparable or
higher than the densities we investigated in our study. From our
results we therefore do not expect a significant impact of dispersal
asymmetry at these density of connections.

The analysis of the results in \citep{Vuilleumier06} is based on the
number of connected patches in contrast to our analysis using the
global mean number of connections $L/N$. The statistics of the number
of connected patches seems to differ significantly between the
asymmetric and the symmetric connectivity matrices investigated, a
phenomenon we were not able to reproduce. In particular the example of
a symmetric random pattern with more than $85$ connections per patch
but only $96$ connected patches raises questions, since the largest
cycle of closed dispersal in non-regular connectivity matrices we
generated always extended to at least $99$ patches for densities above
$7$ connections per patch with a strong trend towards $100$ patches
with increasing density. For this reason we assume, that the effects
described in \citep{Vuilleumier06} originate from differences in
network topology between the investigated connectivity matrices rather
than differences in dispersal asymmetry.

\citet{Bode08} investigated the same metapopulation model as in the
present work in a slightly different setup ($N=10$, $e=0.4$, and
$L/N=2.6$). Instead of simulating individual realisations, the
probability of metapopulations to go extinct within $100$ time steps
was calculated numerically for different dispersal patterns. This
method restricts the analysis to rather small metapopulations of $10$
patches. Extinction probabilities were calculated for metapopulations
connected through different dispersal patterns generated by the small
world algorithm \citep[see e.g.][]{Watts98,Kininmonth09} initiated
with a particular symmetric dispersal pattern (Bode, pers.\
communication).  \citet{Bode08} concluded from qualitative graphical
analysis of their simulation results\footnote{In our point of view a
  correlation between the extinction probability and dispersal
  asymmetry is not obvious from the Figure the authors refer to
  \citep[p.\ 205, Fig.\ 3]{Bode08}.  Bode, however, kindly provided
  additional data on an accordant simulation, which indeed shows a
  negative impact of dispersal asymmetry on the metapopulation
  extinction probability after $100$ time steps.}, that asymmetry
reduces persistence and exhibits a distinct threat to metapopulations.

The discussion of our results in Section
\ref{sec:discussion:interpret} relates our graphical analysis to the
extinction probability in a certain number of time steps\footnote{For
  the parameters marked green within $100$ time units extinctions
  probabilities below $1-\exp(-5.1\times 10^{-8}\times 100)\approx
  5\times 10^{-6}$ are expected, for the red regions an accordant
  calculation yields probabilities above almost $0.5$.}, which allows
for a comparison of the results. From additional simulation data we
received from Bode it seems, that the negative effect in their
approach is larger than what we would expect from our simulation for
the general, non-regular case (Section
\ref{sec:discussion:regul}). Additional simulations performed for
metapopulations likewise subjected to non-regular dispersal patterns
but reduced to the size of $10$ patches indicated a general increase
in the probability of extinction but no significant impact of
metapopulation size on the impact of symmetry. We therefore assume,
that the differences related to symmetry observed by
\citeauthor{Bode08} partly are owed to the fact, that the patterns in
their study were generated from a particular symmetric starting
configuration of the small world algorithm and that the similarity of
patterns to this starting configuration correlates with the symmetry
properties.

Recently another work was devoted to the effect of asymmetry on
metapopulation viability \citep{Vuilleumier10}. This work aims to
cover different aspects of asymmetry simultaneously, which makes it
difficult to ascribe the variety of effects observed to certain
properties of dispersal matrices. One configuration, however, seems to
be equivalent to the simulations we performed for general dispersal
matrices in Section \ref{sec:discussion:regul} for anti-symmetric and
symmetric dispersal, respectively \citep[][p.\ 229, Fig.\ 2, right
column]{Vuilleumier10}. The results the authors obtain on these
patterns are in agreement with our observations, that the degree of
symmetry of dispersal matrices has no significant impact on
metapopulation viability at intermediate density of dispersal
connections \citep[cp.][p.\ 213, Fig.\ 6, difference between the plots
in the right column]{Vuilleumier10}.

\section{\label{sec:conclusions}Conclusions}
We investigated the consequences of the symmetry of dispersal patterns
on the viability of metapopulations. Our analyses are based on
simulations of a stochastic patch occupancy model.

First we define the degree of dispersal symmetry, $\gamma$, which is
based on the symmetry of the connectivity matrix (Equation
\ref{eq:def-deg-asymm}). In order to be able to minimise possibly
confounding effects we restrict our main analysis to regular dispersal
patterns, where asymmetry does neither affect the homogeneity of
dispersal nor the local balances of incoming and outgoing dispersal
connections. For these patterns we do not see any negative effect of
dispersal asymmetry. For the more general case of non-regular
dispersal patterns minor negative effects of asymmetric dispersal on
metapopulation viability are confirmed, but only at rather weak
densities of dispersal (cp.\ Section \ref{sec:discussion:regul}). At
these densities differences in dispersal symmetry generally are
accompanied by other hierarchical differences of the dispersal
network. This e.g.\ becomes evident from a neat example of a two patch
metapopulation investigated in detail in \citep[p.\ 208, Appendix
A]{Bode08}, where dispersal asymmetry by return results in a
source-sink problem.

From first instance it is not self-evident whether these accompanying
effects are the origin or a consequence of asymmetric dispersal, since
their characteristic strongly depends on how the system of study was
constructed and chosen. For realistic dispersal patterns the solution
proposed in \citep{Vuilleumier10}, namely to investigate dispersal
asymmetry independent from the discussion of sources and sinks,
however does not seem to work out, since these effects in general are
strongly connected to one another. These correlations in the past made
the investigation of asymmetric dispersal highly dependent on the
system of study, which was the main difficulty in understanding the
role of dispersal asymmetry.  In order to resolve this problem we
suggest to discuss the symmetry of dispersal patterns at large scales
e.g.\ based on a definition similar to Equation
\eqref{eq:def-deg-asymm} and the statistics of sources and sinks, the
homogeneity of the dispersal network, and other features
characterising the local flow of migrants \emph{jointly} instead of in
isolation.

It was the aim of the present work, to clarify the role of asymmetric
dispersal and its impact on metapopulation viability.  In contrast to
previous studies \citep{Vuilleumier10,Vuilleumier06,Bode08} we see
only weak effects of asymmetric connectivity on metapopulation
extinction, which suggests that natural populations with asymmetric
dispersal may not per se suffer from increased extinction
risks. Instead effects observed in simulations, real world data, or in
the evaluation of management strategies \citep[see e.g.][]{Haight08}
might be reflected more significantly by other features of complex
dispersal patterns. A promising path towards a discussion of
potentially important features is taken in the investigations of the
viability of metapopulations connected through a variety of different
dispersal patterns as provided in \citep{Bode08,Artzy-Randrup10}. We
expect that eventually only a theoretical analysis of the stochastic
metapopulation model applied can reveal the features relevant for
metapopulation viability.

\section{\label{sec:ack}Acknowledgements}
We kindly acknowledge comments by Bernt Wennberg on an early version
of the manuscript and suggestions by Kerstin Johannesson on a more
recent version. We kindly appreciate that Michael Bode contributed
simulation results and shared details on his 2008 work,
\citep{Bode08}.  Furthermore we are deeply indebted to kind and
constructive comments of two anonymous reviewers. This work was
supported by a Linnaeus-grant from the Swedish Research Councils, VR
and Formas (http://www.cemeb.science.gu.se), by FORMAS through
contract 209/2008-1115 (PRJ), and by the Swedish Research Council
through contract 275 621-2008-5456 (PRJ).


\appendix
\section{\label{sec:algor-gener-balanc}Algorithm for the generation of
  regular dispersal patterns}
Since we intended to compare cases primarily differing in their
symmetry properties, we focused on \emph{regular} dispersal patterns
with fixed number of in- and out-going dispersal routes for every
patch. For the connectivity matrices $D$ this is equivalent to the
constraint that the sums over every column and every row are equal,
that is
\begin{equation}
  \sum_{k}d_{ik}=\sum_{k}d_{kj}=L/N\label{eq:l-rew-col}
\end{equation}
for any $i$ and $j$. Here $L$ is the total number of activated
dispersal routes.

Random matrices at arbitrary degree of symmetry that are complying
with Equation \eqref{eq:l-rew-col} are generated by the following
algorithm, that is repeated until a matrix $D$ with $L$ non-zero
elements is obtained:
\begin{enumerate}
\item Set $D=0$, generate a random matrix
  $B\in\left[0,1\right]^{N\times N}$, where $b_{ij}$ are random
  numbers drawn independently from an arbitrary distribution.  For
  instance uniformly distributed random variables are suitable
  here. Ensure that all elements of $B$ are unique.
\item Set diagonal elements $b_{ii}$ to $10$ for all $i$.
\item Calculate the desired number of symmetric connections,
  $n_s=\gamma L$
\item Repeat until smallest element of $B$ is larger
  than $1$ or $\sum_{i,j}d_{ij}=L$:
  \begin{enumerate}
  \item Identify row $i$ and column $j$ of the smallest value of $B$
  \item Set $d_{ij}=1$ and $b_{ij}=10$ ({*})
  \item If $\sum_{k}d_{ik}=L/N$ set $b_{ik}=10$ for every $k$ ({*})
  \item If $\sum_{k}d_{kj}=L/N$ set $b_{kj}=10$ for every $k$ ({*})
  \item Switch $i$ and $j$
  \item if $n_s >0$ (generate symmetric connection):
    \begin{itemize}
    \item repeat the steps marked by ({*})
    \item reduce $n_s$ by $2$
    \end{itemize}
    else: (generate asymmetric connection)
    \begin{itemize}
    \item set $b_{ij}=10$
    \end{itemize}
  \end{enumerate}
\item Reject result if $\sum_{i,j}d_{ij}<L$.
\end{enumerate}
Note that the value $10$, of course, is arbitrary. Any number greater
than $1$ is suitable to ensure that the corresponding elements of $D$
are not selected by the algorithm. This algorithm randomly orders the
elements of $D$ and activates them step by step. It generates random
connectivity matrices with given degree of symmetry and it is
sufficiently efficient for small and intermediate $L$.

The implementation of the algorithm in FORTRAN90 is straightforward
(compilation tested with the GNU Fortran compiler \emph{gfortran
  v4.3.3}): \definecolor{light-gray}{gray}{0.95}%
\lstset{ %
  language=Fortran, %
  basicstyle=\tiny, %
  numbers=left, %
  numberstyle=\tiny, %
  stepnumber=5, %
  numbersep=10pt, %
  backgroundcolor=\color{light-gray}, %
  showspaces=false, %
  showstringspaces=false, %
  showtabs=false, %
  frame=single, %
  tabsize=2, %
  captionpos=t, %
  breaklines=false, %
  breakatwhitespace=false, %
  title=REGULAR.F90, %
}
 
\begin{lstlisting}
PROGRAM REGULAR_CONNECTIVITY
  !======================================================================
  ! PROGRAM REGULAR_CONNECTIVITY
  ! GENERATION OF REGULAR RANDOM DISPERSAL PATTERNS 
  ! TESTED WITH GFORTRAN 4.3.3
  ! (C) 2010 BY DAVID KLEINHANS, UNIVERSITY OF GOTHENBURG, SWEDEN
  ! DISTRIBUTED UNDER THE CREATIVE COMMONS ATTRIBUTION 3.0 LICENSE
  !======================================================================

  IMPLICIT NONE

  INTEGER,PARAMETER::N=100              !METAPOPULATION SIZE
  INTEGER,PARAMETER::MAX_REJECTIONS=1000!MAXIMUM NO OF REJECTED MATRICES

  DOUBLE PRECISION::GAMMA               !DEGREE OF SYMMETRY
  INTEGER::LBYN                         !NO. OF CONNECTIONS PER PATCH

  INTEGER::D(N,N)                       !CONNECTIVITY MATRIX

  DOUBLE PRECISION::RAND(N,N)           !RANDOM MATRIX USED FOR ORDERING OF LINKS
  DOUBLE PRECISION::REMAINING_SYM       !NO OF REMAINING SYMMETRIC CONNECTIONS 
  INTEGER::REJECTIONS                   !COUNT NUMBER OF REJECTED DISPERSAL PATTERNS
  INTEGER::LOC(2)                       !LOCATION OF THE SMALLEST ELEMENT OF RAND
  INTEGER::I,J                          !AUXILIARY VARIABLES, USED FOR LOOPS ONLY
  LOGICAL::GRIDOK                       !CHECK IF GRID COMPLIES WITH CONSTRAINTS


  ! === INITIALIZE RANDOM NUMBER GENERATOR ===
  CALL RANDOM_SEED

  ! === REQUEST PARAMETERS ===
  WRITE(*,"(A,I4)")"REGULAR DISPERSAL MATRIX FOR METAPOPULATION OF SIZE N=",N
  WRITE(*,"(A)")"PLEASE ENTER PARAMETERS:"
  WRITE(*,"(A)")"DEGREE OF SYMMETRY, GAMMA (DOUBLE PRECISION, >=0 AND <=1)?"
  READ(*,*)GAMMA
  WRITE(*,"(F8.5)")GAMMA
  WRITE(*,"(A,I4,A)")"NO OF CONNECTIONS PER PATCH, LBYN (INTEGER, >0 AND <",&
       &(N-1)/2,")?"
  READ(*,*)LBYN
  WRITE(*,"(I4)")

  ! === GENERATE DISPERSAL PATTERN ===
  GRIDOK=.FALSE.
  REJECTIONS=0
  DO WHILE(.NOT.GRIDOK) 

     ! == STARTING CONFIGURATION: ==
     !    ALL LINKS INACTIVE
     D=0
     !    CALCULATE NUMBER OF SYMMETRIC LINKS TO BE GENERATED
     REMAINING_SYM=NINT(GAMMA*LBYN*N)
     !    GENERATE RANDOM NUMBER MATRIX FOR ORDERING OF LINKS
     !    (EXCLUDE DIAGONAL ELEMENTS BY ASSIGNING VALUE OF 10)
     DO I=1,N
        DO J=1,N
           IF(I.NE.J)THEN
              CALL RANDOM_NUMBER(RAND(I,J))
           ELSE
              RAND(I,J)=10.D0
           ENDIF
        ENDDO
     ENDDO

     ! == ADD CONNECTIONS UNTIL NO LINKS ARE AVAILABLE ANY MORE ==
     DO WHILE((MINVAL(RAND).LT.1).AND.(COUNT(D.EQ.1).LT.LBYN*N))

        !LOCATE THE SMALLEST ELEMENT OF RAND
        LOC=MINLOC(RAND)

        !SET RANDOM NUMBER OF THE ELEMENT TO 10 AND ACTIVATE CORRESPONDING LINK 
        RAND(LOC(1),LOC(2))=10.D0
        D(LOC(1),LOC(2))=1
        !CHECK WHETHER NUMBER OF DESIRED INCOMING OR OUTGOING LINKS ALREADY
        !HAS BEEN REACHED FOR THE PATCH OF FOCUS, PREVENT FURTHER LINKS IF SO
        IF(COUNT(D(LOC(1),:).EQ.1).GE.LBYN)RAND(LOC(1),:)=10.D0
        IF(COUNT(D(:,LOC(2)).EQ.1).GE.LBYN)RAND(:,LOC(2))=10.D0

        !IF SYMMETRIC CONNECTIONS ARE REMAINING: MAKE THE CURRENT A SYMMETRIC ONE,
        !ELSE ENSURE THAT THE REVERSE DIRECTION IS NOT ACTIVATED
        IF(REMAINING_SYM.GT.0)THEN
           D(LOC(2),LOC(1))=1
           RAND(LOC(2),LOC(1))=10.D0
           IF(COUNT(D(LOC(2),:).EQ.1).GE.LBYN)RAND(LOC(2),:)=10.D0
           IF(COUNT(D(:,LOC(1)).EQ.1).GE.LBYN)RAND(:,LOC(1))=10.D0
           REMAINING_SYM=REMAINING_SYM-2
        ELSE
           RAND(LOC(2),LOC(1))=10.
        ENDIF
     ENDDO

     !CHECK WHETHER THE DESIRED NO OF LINKS HAS BEEN GENERATED
     !REJECT AND RESTART IF NOT, ACCEPT THE PATTERN OTHERWISE
     IF (COUNT(D.EQ.1).EQ.LBYN*N)THEN
        GRIDOK=.TRUE.
     ELSE
        REJECTIONS=REJECTIONS+1
        IF(REJECTIONS.LT.MAX_REJECTIONS)THEN
           WRITE(*,"(A,I4,A)")"PATTERN ",REJECTIONS,&
                &" REJECTED, RESTARTING GRID GENERATION ..."
        ELSE
           WRITE(*,"(A)")"GRID GENERATION NOT SUCCESSFULL."
           WRITE(*,"(A)")"PLEASE TRY LOWER LBYN OR INCREASE MAX_REJECTIONS."
           STOP
        ENDIF
     ENDIF
  ENDDO

  ! === WRITE D TO STANDARD OUTPUT ===
  WRITE(*,*)"CONNECTIVITY MATRIX D:"
  DO I=1,N
     WRITE(*,"(999I1.1)")D(I,:)
  ENDDO
  WRITE(*,"(A)")"DONE!"

END PROGRAM REGULAR_CONNECTIVITY
\end{lstlisting}

\end{document}